\documentclass[conference]{IEEEtran}
\usepackage{cite}
\usepackage{amsmath,amssymb,amsfonts}
\usepackage{algorithmic}
\usepackage{algorithm}
\usepackage{graphicx}
\usepackage{textcomp}
\usepackage{xcolor}
\usepackage{tabularx}
\usepackage{bbm}
\usepackage{multirow}

\begin{document}

\title{Phasic Policy Gradient Based Resource Allocation for Industrial Internet of Things \\
}

\author{\IEEEauthorblockN{Lokesh Bommisetty\IEEEauthorrefmark{1} and
Venkatesh Tiruchirai Gopalakrishnan\IEEEauthorrefmark{2}}
\IEEEauthorblockA{Department of Electrical Engineering, Indian Institute of Technology Madras, Chennai, India - 600036}
%\IEEEauthorblockA{(ee18d701@smail.iitm.ac.in\IEEEauthorrefmark{1},tgvenky@ee.iitm.ac.in\IEEEauthorrefmark{2})}
}
\maketitle

\begin{abstract}

 Time Slotted Channel Hopping (TSCH) behavioural mode has been introduced in IEEE 802.15.4e standard to address the ultra-high reliability and ultra-low power communication requirements of Industrial Internet of Things (IIoT) networks. Scheduling the packet transmissions in IIoT networks is a difficult task owing to the limited resources and dynamic topology. In this paper, we propose a phasic policy gradient (PPG) based TSCH schedule learning algorithm. The proposed PPG based scheduling algorithm overcomes the drawbacks of totally distributed and totally centralized deep reinforcement learning-based scheduling algorithms by employing the actor-critic policy gradient method that learns the scheduling algorithm in two phases, namely policy phase and auxiliary phase. 
 %In this method, we show that the schedule converges quickly compared to any other actor-critic method and also improves the system throughput performance by $58\%$ compared to the minimal scheduling function, a default TSCH schedule.
\end{abstract}


\section{Introduction}
Industrial Internet of Things (IIoT) has revolutionised the manufacturing processes due to its low-cost solutions for massive data collection features \cite{longo2018optimal}. The devices in IIoT are mostly battery operated and hence cannot afford to keep their radio on for long hours. Hence the transmission, reception and sleep routines of nodes has to be scheduled to save their energy and to provide a synchronized connection in the network. The nodes in TSCH network communicate using a scheduling matrix composed of cells which are designated by its time slot index and the channel index. Recently, learning algorithms such as machine learning (ML) and reinforcement learning (RL) methods are being adopted to address resource allocation problems in wireless networks \cite{cunha2021intelligent}. Cobbe \textit{et al.} proposed the PPG, by modifying the traditional actor-critic  policy gradient method \cite{cobbe2021phasic}. In PPG, the policy and the value functions are learnt independently so that the sharing of network parameters between the value optimization and the policy optimization networks can be regulated. Hence PPG overcomes the disadvantages of sharing network parameters in the traditional actor-critic policy gradient methods.

%Recently, learning algorithms such as machine learning (ML) and reinforcement learning (RL) methods are being adopted to address resource allocation problems in wireless networks \cite{cunha2021intelligent}. Policy gradient methods of reinforcement learning target at modeling and optimizing the policy directly by using a parameterizing the policy, where the policy is the probability distribution of action taken by a node at a given time. Schulman \textit{et al.} proposed a proximal policy optimization algorithm, a policy gradient method, by clipping the objective function to limit the divergence of parameterized policy learning \cite{schulman2017proximal}. Cobbe \textit{et al.} proposed the PPG, by modifying the actor-critic  policy gradient method \cite{cobbe2021phasic}. In PPG, the policy and the value functions are learnt independently so that the sharing of network parameters between the value optimization and the policy optimization networks can be regulated. Hence PPG overcomes the disadvantages of sharing network parameters in the traditional actor-critic policy gradient methods.

\section{Problem Formulation}
\label{sec:problem}
We consider the TSCH network consisting of a set of $N$ nodes communicating to a single border router that collects data from the nodes in multiple hops. 
%The routing protocol for low power lossy networks (RPL) is employed to create a routing topology among the nodes. 
In TSCH, time is discretized into slots with time-slot index $t=0,1,2,\dots$. The communication bandwidth is divided into $M$ number of channels indexed by $k$, where $k=1,2,\dots,M$. Let us define a indicator random variable $X^{i}_{(t,k)}$ that takes value $1$ if node $i$ is scheduled in the cell $(t,k)$ and takes $0$ otherwise. Let $T$ number of timeslots together constitute a slotframe \cite{molisch2004ieee}.

Let us consider $D_i$ to be the maximum delay that can be experienced by a packet generated by node $i$. Packets will be dropped if the waiting time of the packet in queue ($W$), is more than the deadline $D_i$. Let $P^i_{k,drop}=Pb(W>D_i)$ be the probability of a packet generated by node $i$ being dropped in the network due to the violation of deadline constraint. Let $P^i_{k,err}= Pb(\Gamma^i_k<\beta)$ be the probability that a packet is discarded due to co-channel interference where $\Gamma^i_k$ is the instantaneous signal to interference and noise ratio (SINR) of node $i$ transmitting on channel $k$ and $\beta$ is the threshold. With the above given cases of packet failure, the success probability of a packet generated by node $i$ and transmitted on channel $k$ is given as $P^i_{k,suc}=1-P^i_{k,drop}-P^i_{k,err}+P^i_{k,drop}P^i_{k,err}$. The network throughput can be determined as follows.

\begin{equation}
    Th= \sum_{i=1}^{N}\sum_{k=1}^M P^i_{k,suc}
\end{equation}

 The energy efficiency of the network is given by the transmission rate achieved by the network per unit transmission power spent. Considering $p^i$ to be the transmitting power of node $i$ the energy efficiency can be defined as follows

\begin{equation}
    \eta = \frac{\sum_{i=1}^{N}\sum_{k=1}^M P^i_{k,suc}}{\sum_{i=1}^{N}\sum_{k=1}^Mp^iX^i_{(t,k)}}
\end{equation}
where $p^i$ is the transmitting power of node $i$.

 The resource allocation is given by the schedule $\textbf{X}=(X^i_{(t,k)})_{i\in \mathcal{N}, k \in \{1,2,\dots,M\},t \in \{1,2,\dots,T\}}$ and the power allocation vector $\textbf{P}=(p_i)_{i\in \mathcal{N}}$. In our paper, the objective is to jointly optimize the cell allocation and power allocation to nodes in the network to maximise the network throughput and energy efficiency while guaranteeing the QoS requirements. Hence, the resource allocation problem can be formulated as an optimization problem as follows. 

\begin{equation}
    \underset{\textbf{P},\textbf{X},\theta_1,\theta_2}{max}\qquad U=\theta_1 Th + \theta_2 \eta
    \label{Utility}
\end{equation}
\begin{equation}
    \text{s.t.}\qquad \sum_{k=1}^M X^{i}_{(t,k)}\leq 1, \qquad \forall i \in \mathcal{N} \qquad\qquad
    \label{C1}
\end{equation}
\begin{equation}
    P^i_{k,drop}<\epsilon_i \qquad \forall i \in \mathcal{N}, k \in \{1,2,\dots,M\}
    \label{C2}
\end{equation}
\begin{equation}
    P^i_{k,err}<\delta_i \qquad \forall i \in \mathcal{N},k \in \{1,2,\dots,M\}
    \label{C3}
\end{equation}

where $\theta_1$ and $\theta_2$ are the optimization parameters giving weightage to throughput and energy efficiency respectively. The maximisation of the objective function as shown in \eqref{Utility} is subjected to the constraints given in equations \eqref{C1},\eqref{C2} and \eqref{C3}. The constraint in \eqref{C1} tells that a node cannot transmit on more than one channel in a given time-slot. Equations \eqref{C2} and \eqref{C3} ensures the QoS requirements of the network in terms of the deadline based delivery and error probability respectively. 
%The above optimization problem is a non-linear convex optimization problem and is difficult to solve in polynomial time. Further, the domain of the problem is a mixture of continuous and binary domains since the scheduling matrix $\textbf{X}$ is a high dimension binary matrix and the rest of the variables ($\textbf{P}, \theta_1$ and $\theta_2$) are continuous variables.
\section{PPG for TSCH resource allocation}
\label{main}
In this section, we discuss the usage of PPG method to solve the above formulated optimization problem for resource allocation for TSCH networks. The state space of the optimization problem is defined by $\mathcal{S}=\{\theta_1,\theta_2,\mathbf{Q},\mathcal{T}\}$ where $\mathbf{Q}$ is the set of possible queue length vectors and $\mathcal{T}$ is the set of network topologies. Action space $\mathcal{A}$ is the node's choice of cells and the transmission power \textit{i.e.,} $\{(X_{(t,k)}^i,p^i)_{i=1}^n\} \in \mathcal{A}$. Reward obtained at each step is nothing but the utility function $U$. 
%In subsection \ref{MDP}, we model the proposed optimization problem as a markov decision process (MDP) by defining the three key elements of MDP namely, state, action and reward. In subsection \ref{prilim}, we define and discuss the preliminary quantities required for PPG. Finally, in subsection \ref{PPG}, we adapt PPG method to solve the resource allocation problem formulated in section \ref{sec:problem}.
In PPG, the optimization of the objective function occurs in two phases namely, policy phase and the auxiliary phase. During policy phase, the agent is trained using proximal policy optimization (PPO) \cite{schulman2017proximal}. During auxiliary phase, the features from the value function are distilled into the policy network, so that the future policy phases improve. 

\subsubsection{Policy Phase}
During policy phase, we update the policy network by optimizing the following objective function. 

\begin{equation}
\begin{aligned}
    &\mathcal{L}^{policy}(\theta)=-\nu D_{KL}(\pi_{\theta_{old}}\|\pi_\theta)+\\&{\mathbb{E}_{\pi_{\theta_{old}}}}\left[min\left(x_t(\theta)\hat{A}_{(s_t,\overline{a_t})}^{\pi_{\theta_{old}}},clip\left(x_t(\theta),1-\zeta,1+\zeta\right)\hat{A}_{(s_t,\overline{a_t})}^{\pi_{\theta_{old}}}\right)\right]
\end{aligned}
\end{equation}
In the above equation, $x_t(\theta)=\frac{\pi_\theta(\overline{a_t}|s_t)}{\pi_{\theta_{old}}(\overline{a_t}|s_t)}$ is the ratio of the target policy and the old policy, where $s_t \in \mathcal{S}$ and $a_t \in \mathcal{A}$. $\hat{A}_{(s_t,\overline{a_t})}^{\pi_{\theta_{old}}}$ is the advantage estimator function at time $t$ \cite{schulman2017proximal}. The function $D_{KL}$ is the Kullback-Leibler divergence function that gives the distance between two probability distributions or also known as their relative entropy. Similar to the policy network, we train the value function network by optimizing the following function.

\begin{equation}
    \mathcal{L}^{value}(\phi)=\mathbb{E}\left[\frac{\|V_{\phi}(s_t)-\hat{V}(s_t)\|^2}{2}\right]
\end{equation}

where $\phi$ is the training parameter for the value function analogous to $\theta$ in the policy network. Here, $V_\phi$ and $\hat{V}$ are the target value function and estimated value function respectively.
\subsubsection{Auxiliary Phase}
In auxiliary phase, the joint objective function that includes behavioral cloning loss and an arbitrary auxiliary loss is used to optimized the policy network. 
\begin{equation}
    \mathcal{L}^{joint}=\mathcal{L}^{aux}+\mu_{clone}D_{KL}(\pi_{\theta_{old}}\|\pi_\theta)
\end{equation}

where $\pi_{\theta_{old}}$ is the policy right after the policy phase and just before the beginning of auxiliary phase. Here, if the auxiliary objective is not present, then the optimization just preserves the original policy with hyper-parameter $\mu_{clone}$ regulating the trade-off. The auxiliary function $\mathcal{L}^{aux}$ can be any objective function. In our case, we use the value optimization function as the auxiliary objective as done by Cobbe \textit{et al.} in their paper proposing PPG.

%Algorithm \ref{algo} gives the summary of the above discussed PPG based resource allocation. As explained above, the algorithm consists of policy phase and auxiliary phase. In policy phase, the policy network is updated to find an optimal $\theta$ and the value network is updated to find the value estimation parameter $\phi$. The number of policy update iterations to be done in policy phase is given by $N_\pi$. Number of training epochs performed across the data in buffer $B$ for the policy network and the value network are given by $E_{\pi}$ and $E_V$ respectively. Similarly, the number of training epochs on data in $B$ in auxiliary phase is given by $E_{aux}$.

\section{Simulation Results}
\label{res}
\begin{figure}[t]
    \centering
    \includegraphics[width=\linewidth, height=4cm, trim={36mm 2mm 40mm 16mm},clip]{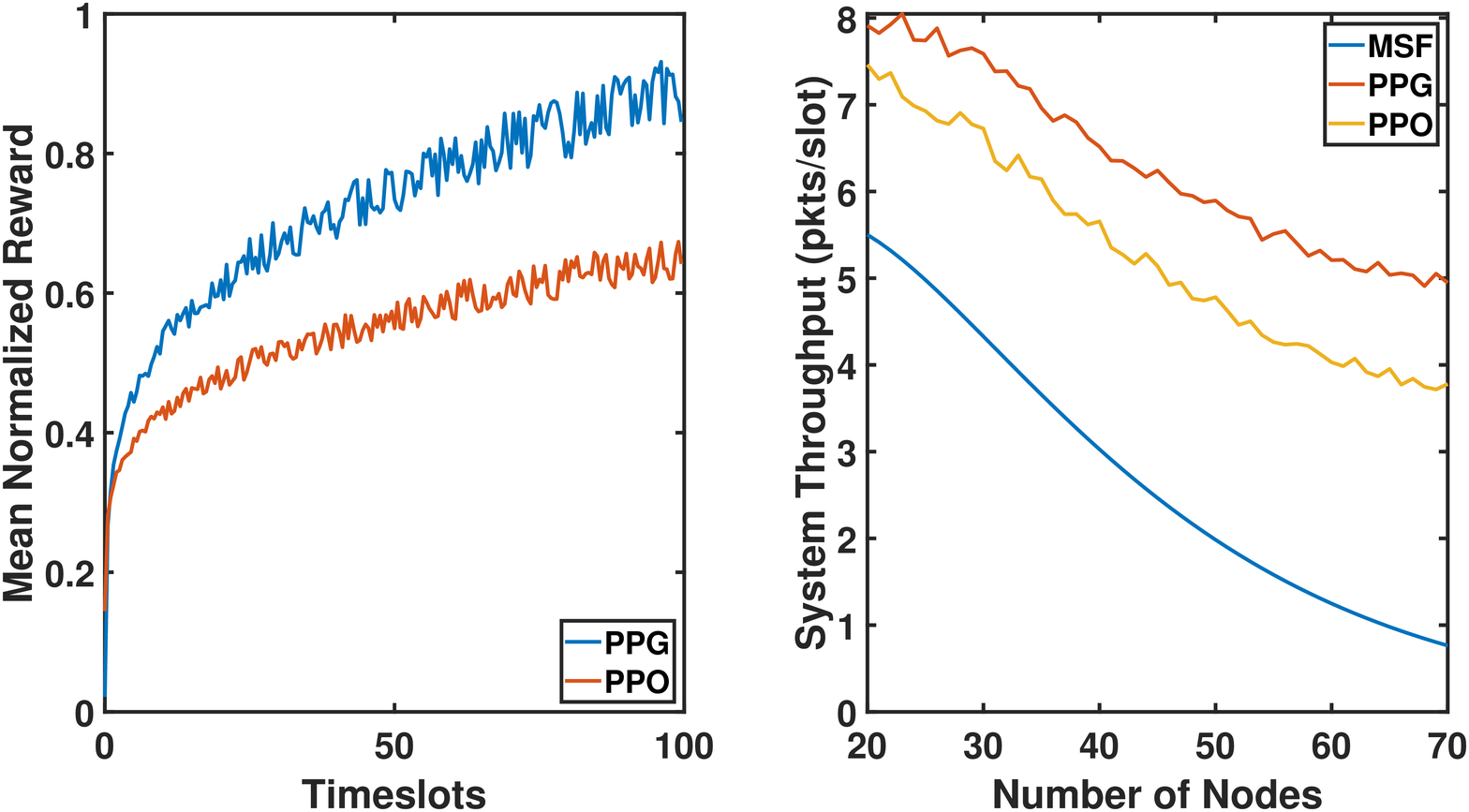}
    \caption{(left) Convergence Performance of PPG and PPO learning algorithms for TSCH schedule. (right) System Throughput performance against the number of nodes in the network.}
    \label{fig:comb}
    \vspace{-4mm}
\end{figure}
To evaluate the performance of PPG based scheduling algorithm in TSCH networks, we have simulated the network using 6TiSCH simulator. We consider that the network consists of 70 nodes communicating to a border router over a maximum of 3 hops. We compare the performance of our proposed scheduling algorithm with that of PPO \cite{schulman2017proximal} based algorithm and the Minimum Scheduling Function (MSF) \cite{chang20196tisch}, which is a default scheduling algorithm in TSCH networks.

Firstly, we discuss the convergence performance of the proposed PPG based TSCH scheduling algorithm in comparison with PPO learning algorithm in Fig. \ref{fig:comb} (left). It can be seen that the PPG achieves a better reward than PPO right from the starting of the training phase and converges quickly. We show the performance of the network in terms of the system throughput when different scheduling schemes are implemented. In Fig. \ref{fig:comb} (right), we show the throughput performance when PPG and PPO based scheduling algorithms are implemented in comparison with the default scheduling algorithm MSF. As the number of nodes in the network increase, the interference on each link increases resulting in the increase of error probability of transmitted packets and thus reducing the system throughput. We show that the PPG improves the system throughput by $58\%$ compared to the default scheduling algorithm MSF and $22\%$ over the proximal policy optimization method (PPO). As a part of future work, we wish to explore auxiliary functions to account for the packet drops by employing a penalty in the policy network to further improve the convergence performance.

\end{document}